\documentclass[twocolumn,amsmath,amssymb,showpacs]{revtex4}
\usepackage{bm}
\usepackage{graphicx}
\usepackage{epstopdf}

\begin{document}

\newcommand{\uu}[1]{\underline{#1}}
\newcommand{\pp}[1]{\phantom{#1}}
\newcommand{\be}{\begin{eqnarray}}
\newcommand{\ee}{\end{eqnarray}}
\newcommand{\ve}{\varepsilon}
\newcommand{\vs}{\varsigma}
\newcommand{\Tr}{{\,\rm Tr\,}}
\newcommand{\pol}{\frac{1}{2}}
\newcommand{\RR}{\rotatebox[origin=c]{180}{$\mathbb{R}$} }
\newcommand{\CC}{\rotatebox[origin=c]{180}{$\mathbb{C}$} }
\newcommand{\rr}{\mathbb{R}}
\newcommand{\Exp}{{\,\rm Exp\,}}
\newcommand{\Sin}{{\,\rm Sin\,}}
\newcommand{\Cos}{{\,\rm Cos\,}}
\newcommand{\Sinh}{{\,\rm Sinh\,}}
\newcommand{\Cosh}{{\,\rm Cosh\,}}

\title{
If gravity is geometry, is dark energy just arithmetic?}
\author{Marek Czachor$^{1,2}$}
\affiliation{
$^1$Katedra Fizyki Teoretycznej i Informatyki Kwantowej,
Politechnika Gda\'nska, 80-233 Gda\'nsk, Poland,\\
$^2$Centrum Leo Apostel (CLEA),
Vrije Universiteit Brussel, 1050 Brussels, Belgium,\\
}

\begin{abstract}
Arithmetic operations (addition, subtraction, multiplication, division), as well as the calculus they imply, are non-unique. The examples of four-dimensional spaces, $\mathbb{R}_+^4$ and $(-L/2,L/2)^4$, are considered where different types of arithmetic and calculus coexist simultaneously. In all the examples there exists a non-Diophantine arithmetic that makes the space globally Minkowskian, and thus the laws of physics are formulated in terms of the corresponding calculus. However, when one switches to the `natural' Diophantine arithmetic and calculus, the Minkowskian character of the space is lost and what one effectively obtains is a Lorentzian manifold. I discuss in more detail the problem of electromagnetic fields produced by a pointlike charge. The solution has the standard form when expressed in terms of the non-Diophantine formalism. When the `natural' formalsm is used, the same solution looks as if the fields were created by a charge located in an expanding universe, with nontrivially accelerating expansion. The effect is clearly visible also in solutions of the Friedman equation with vanishing cosmological constant. All of this suggests that phenomena attributed to dark energy may be a manifestation of a miss-match between the arithmetic employed in mathematical modeling, and the one occurring at the level of natural laws. Arithmetic is as physical as geometry.
\end{abstract}
\maketitle

\section{Introduction}

The idea of relativity of arithmetic follows from the observation that the four basic arithmetic operations (addition, subtraction, multiplication, division) are fundamentally non-unique, even if one assumes commutativity and associativity of `plus' and `times', and distributivity of `times' with respect to `plus'. The ambiguity extends to calculus and algebra since even the most elementary notions, such as derivatives or matrix products, involve arithmetic operations, sometimes accompanied by limits (`to zero', say). A `zero', the neutral element of addition, inherits its ambiguity from the ambiguity of addition. The same concerns a `one', the neutral element of multiplication.

The freedom of choosing arithmetic and its corresponding calculus is a universal symmetry of any mathematical model, but we are still lacking its physical understanding. For all that, treated just as a mathematical trick, the idea has found concrete applications in fractal theory \cite{MC,ACK,ACK1,ACK2}.

In the paper, I discuss a situation where there is a miss-match between the arithmetic employed at the level of mathematical principles, and the one that is naturally `employed' by Nature. We will see that consequences of the miss-match can be similar to those of dark energy.

In natural sciences there exists at least one example of such a miss-match, and we experience it in our everyday life.  This is the Weber phenomenon known from neuroscience \cite{BairdNoma,Norwich}. Namely, it is an experimental fact that the increment $x\to x+ kx$ of intensity of a stimulus (sound, light, taste, etc.) is perceived by our nervous system as being independent of $x$. $k$ depends on the type of stimulus and is known as a Weber constant. The Weber law
$\Delta x/x=k\approx\textrm{const}$ is valid in a wide range of stimulus parameters.

From a mathematical point of view the Weber law appears as the solution of the following problem \cite{MC1}: Find a generalized subtraction $\ominus$ such that
\be
(x+kx)\ominus x
=
f^{-1}\big(
f(x+kx)-f(x)
\big)
=
\delta x
\ee
is independent of $x$. The solution $f$ is unique and is given by a logarithm, as shown by G.~Fechner in 1850 \cite{Fechner}. This is the reason why decibels correspond to a logarithmic scale. We hear, see, taste and feel the world outside of us though a logarithmic channel of our neurons, although we are typically as unaware of it, as we are unaware of experiencing curvature of space when we feel our weight.

The Fechner problem extends to any natural science.
In order to appreciate it, consider the function $f(x)=x^3$, and define
\be
x\oplus y &=& f^{-1}\big(f(x)+f(y)\big)=\sqrt[3]{x^3+y^3},\label{+}\\
x\ominus y &=& f^{-1}\big(f(x)-f(y)\big)=\sqrt[3]{x^3-y^3},\label{-}\\
x\odot y &=& f^{-1}\big(f(x)f(y)\big)=xy,\label{.}\\
x\oslash y &=& f^{-1}\big(f(x)/f(y)\big)=x/y.\label{/}
\ee
Here multiplication is unchanged.

Mathematicians would generally say that we have introduced two types of {\it field\/}, related by the field isomorphism $f$, but I prefer the terminology of Burgin \cite{Burgin,Burgin1} where formulas such as (\ref{+})-(\ref{/}) occur in arithmetic contexts. A terminology involving `fields' or `relativity of fields' would be very confusing in the context of physics, especially in relativistic field theory. An arithmetic involving a non-trivial $f$ is termed by Burgin a non-Diophantine one, as opposed to the Diophantine case of $f(x)=x$. I will also stick to this distinction, although one should bear in mind that it is largely a matter of convention which of the two arithmetics is Diophantine.

The non-Diophantine derivative
\be
\frac{D A(x)}{Dx}
&=&
\lim_{h\to 0}\Big(A(x\oplus h)\ominus A(x)\Big)\oslash h\label{der0}
\ee
satisfies all the basic rules of differentiation (the Leibnitz rule for $\odot$, the chain rule for composition of functions, linearity with respect to $\oplus$...).
The solution of
\be
\frac{D A(x)}{Dx}=A(x),\quad A(0)=1,\label{exp eq}
\ee
is unique, but it comes as some surprise, at least when one first encounters it, that
\be
A(x)=e^{x^3/3},\label{exp sol}
\ee
as one can verify directly from definition (\ref{der0}).
Had one replaced $f(x)=x^3$ by $f(x)=x^5$, one would have found
$A(x)=e^{x^5/5}=f^{-1}(e^{f(x)})$, as the reader has probably already guessed.

Notice that the change of $f$ is not a change of variables. The differential equation (\ref{exp eq}) remains linear in spite of non-linearity of $f$. The change of $f$ cannot be regarded as a change of gauge either, with covariant derivative $D/Dx$, since the corresponding connection would be trivial, in spite of non-triviality of $f$. The change of arithmetic operates at a more primitive level than a change of variables or gauge.

Of course, one can denote $g=f^{-1}$ and rewrite (\ref{+})-(\ref{-}) as
\be
x+y &=& g^{-1}\big(g(x)\oplus g(y)\big) = \left(\sqrt[3]{x}\oplus\sqrt[3]{y}\right)^3,\label{+'}\\
x-y &=& g^{-1}\big(g(x)\ominus g(y)\big) = \left(\sqrt[3]{x}\ominus\sqrt[3]{y}\right)^3.\label{-'}
\ee

Now comes the fundamental question: Which of the two additions, $+$ or $\oplus$, is more natural or physical? Which of them is Diophantine, and which is not? Clearly, $\pm$ and $\oplus$, $\ominus$ coexist in the same set, and it is hard to say why (\ref{+})-(\ref{-}) should be regarded as less simple, or more weird, than (\ref{+'})-(\ref{-'}).

Put another way, what kind of a rule is responsible for the implicit preference of $f(x)=x$ over any other one-to-one $f$? The Ockham razor?

Expressing it yet differently, let us assume that it is indeed the `natural' (whatever it means) arithmetic with $\pm$ etc. that we should employ in practical computations. What kind of a rule guarantees that the laws of physics (variational principles, say) are formulated in terms of the same arithmetic and calculus, and not by means of $\oplus$, $\odot$, $D/Dx$, and the like, for some unknown $f$?

I believe the questions are open.

It is not completely unrelated to mention that the case of a linear $f$ was extensively studied by Benioff \cite{Benioff} in a somewhat different context. Here, the departure point was the observation that natural numbers $\mathbb{N}$ can be identified with any countable well-ordered set whose first two elements define 0 and 1. For example, the set $\mathbb{N}_2=\{0,2,4,\dots\}$ of even natural numbers may be regarded as a representation of $\mathbb{N}$. The {\it value function\/} $f(x)=x/2$ maps a natural number $x\in \mathbb{N}_2$ into its value, but to make the structure self-consistent one has to redefine the multiplication: $x\odot y=f^{-1}\big(f(x)f(y)\big)=xy/2$. So, in Benioff's approach numbers themselves are just elements of some formal axiomatic structure, while their values are determined by appropriate linear value maps, which are simultaneously used to redefine the arithmetic operations. The idea was then extended by Benioff to real and complex numbers, and generalized in many ways, including space-time dependent fields of value functions \cite{Benioff1,Benioff2}.

From the Burgin perspective it is essential that one can also encounter nonlinear functions $f$ (Fechner's logarithms, Cantor-type functions for Cantor sets \cite{MC,ACK}, Peano-type space-filling curves in Sierpi\'nski-set cases \cite{ACK1}...). Whenever one tries to apply a Burgin-type generalization to a physical system, one immediately encounters the problem that nontrivial $f$s typically require dimensionless arguments in $f(x)$. Still, physical $x$s are dimensional. One has to associate with a physically meaningful $x$ a dimensionless number, and we return to the problem described by Benioff. Indeed, a dimensionless $x$ is obtained for the price of including a physical unit (of length, say), and units can be chosen arbitrarily. The choice of units effectively introduces a value map, and a change of scale changes this map. Burgin's $f$, when employed in a way proposed in \cite{MC}, can be an arbitrary bijection, so can be composed with any value map with no loss of bijectivity of the composition. Benioff's value maps are thus intrinsically related with Burgin's $f$s, but are not necessarily equivalent to them. The problem of dimensional vs. dimensionless $x$ was discussed in more detail in \cite{ACK}.

%XXXXXXXXXX

The goal of the present paper is to illustrate these problems on concrete examples from relativistic physics, and to contemplate the possibility of detecting a non-trivial $f$  by means of physical observations. We will see that phenomena of dark-energy variety may suggest the presence of some $f$ between the universe of our mathematical formulas, and the physical Universe.

We will consider two spaces, $\mathbb{R}_+^4$ and $(-L/2,L/2)^4$, which become Minkowski space-times when appropriate arithmetic is selected. We will then consider the problem of electromagnetic fields produced by pointlike sources, but the Maxwell field will be formulated in terms of this concrete arithmetic $\oplus$, $\ominus$, $\odot$, $\oslash$, which will make the space Minkowskian. However, when we switch back to the `standard' arithmetic the geometry becomes locally Lorenzian, with a non-Minkowskian global structure. Static charges will appear moving toward event horizons of the space, thus creating an impression of an expanding universe, with accelerating expansion. All of this happens in spite of the fact that the spaces are static, no matter which arithmetic one works with.

The paper is organized as follows. In the next section, two types of real numbers, equipped with their own arithmetic and calculus, are introduced. The two types of reals are related by a bijection $f$. In Section~III, the two types of reals are employed in construction of two types of Minkowski spaces.
Sections IV and V discuss in detail the two concrete examples of Minkowski spaces: $\mathbb{R}_+^4$ and $(-L/2,L/2)^4$. Section VI discusses electromagnetic fields produced by a pointlike charge, with Maxwell equations formulated in terms of a non-Diophantine formalism defined by some $f$. The formalism is implicitly non-Diophantine since an observer who employs in his observations and calculations the arithmetic defined by the same $f$ will never discover that some nontrivial $f$ is implicitly involved. However, in case there is a miss-match, i.e. two different $f$s come into play, the conflict of arithmetics can have observational consequences. This is discussed on explicit examples in sections VII and VIII.
Finally, in section IX, an example of a non-Diophantine-calculus Friedman equation is discussed. We observe accelerated expansion with vanishing cosmological constant.

\section{`Lower' and `upper' reality}

Consider real numbers $\mathbb{R}$ equipped with the `standard' arithmetic operations of addition ($+$), subtraction ($-$), multiplication ($\cdot$), and division ($/$). Let us term these real numbers the `lower reals', and denote them by lowercase symbols. As usual, `$\cdot$' can be skipped: $a\cdot b=ab$. Neutral elements of addition and multiplication in $\rr$ are denoted by 0 and 1.

Now, let us assume that there exist some `upper reals', whose set is denoted by $\RR$. By assumption, \RR is related to $\rr$ by some bijection, $f: \RR\to\mathbb{R}$. Those upper reals are equipped with their own arithmetic and calculus. The arithmetic operations in \RR will be denoted by
 $\oplus$, $\ominus$, $\odot$, $\oslash$. We use the convention that elements of \RR are denoted by upper-case fonts.

The arithmetic in \RR is non-Diophantine in the sense of Burgin \cite{Burgin,Burgin1},
\be
X\oplus Y
&=&
f^{-1}\big(f(X)+f(Y)\big),\\
X\ominus Y
&=&
f^{-1}\big(f(X)-f(Y)\big),\\
X\odot Y
&=&
f^{-1}\big(f(X)f(X)\big),\\
X\oslash Y
&=&
f^{-1}\big(f(X)/f(Y)\big).
\ee
Both arithmetics are commutative, associative, and multiplications are distributive with respect to (appropriate) additions.
Neutral elements of addition and multiplication in \RR are denoted by $0'$ and $1'$, which  implies $0'=f^{-1}(0)$, $1'=f^{-1}(1)$.

A negative in \RR is defined by $\ominus X=0'\ominus X=f^{-1}\big(-f(X)\big)$. Multiplication by zero yields zero in both arithmetics, in particular
$0'\odot X=0'$.

Multiplication is equivalent to repeated addition in the following sense. Let $N\in\mathbb{N}$ and $X'=f^{-1}(X)\in \RR$. Then
\be
N'\oplus X'
&=&
(N+X)',\\
N'\odot X'
&=&
(NX)'\\
&=&
\underbrace{X'\oplus\dots \oplus X'}_{N\rm{ times}}.
\ee
A power function $A(X)=X\odot\dots \odot X$ ($N$ times) will be denoted by $X^{N'}$, since
\be
X^{N'}\odot X^{M'}=X^{(N+M)'}=X^{N'\oplus M'}.
\ee

A derivative of a function $A:\RR\to\RR$ is defined by
\be
\frac{DA(X)}{DX}
&=&
\lim_{H\to 0'}\big(A(X\oplus H)\ominus A(X)\big)\oslash H,\label{DA/DX}
\ee
as contrasted with the derivative of a function $a:\mathbb{R}\to \mathbb{R}$, defined with respect to the lowercase arithmetic,
\be
\frac{da(x)}{dx}=\lim_{h\to 0}\big(a(x+h)-a(x)\big)/h.
\ee
Now let $A=f^{-1}\circ a\circ f$. Then,
\be
\frac{DA(X)}{DX}
&=&
f^{-1}\left(\frac{da\big(f(X)\big)}{df(X)}\right),\label{der}\\
\int_X^YA(X')DX'
&=&
f^{-1}\left(\int_{f(X)}^{f(Y)}a(x)dx\right),\label{int}
\ee
satisfy
\be
\frac{D}{DX}\int_{Y}^X A(X')DX' &=& A(X),\label{calc1}\\
\int_Y^X \frac{DA(X')}{DX'}DX' &=& A(X)\ominus A(Y).\label{calc2}
\ee
Formula (\ref{der}) follows directly from the definitions of $D/DX$ and $d/dx$.
As stressed in the introduction, (\ref{der}) is {\it not\/} the usual formula relating derivatives of $A=f^{-1}\circ a\circ f$ with those of $a$. Indeed, \be
\frac{DA}{DX}  &=& f^{-1}\circ \frac{da}{dx} \circ f,\label{der'}
\ee
so that $D/DX$ behaves like a covariant derivative, but with a trivial connection for {\it any\/} bijection $f: \RR\to \mathbb{R}$. The standard approach, employed in differential geometry or gauge theories, would employ the arithmetic of $\mathbb{R}$, and one would have to assume differentiability of $f$ and $f^{-1}$. Here bijectivity is enough since no derivatives of either $f$ or $f^{-1}$ will occur in (\ref{der}) and (\ref{der'}). This is why this type of calculus is so useful and natural in fractal theory \cite{MC,ACK,ACK1,ACK2}.

%YYYYYYYYYYYYYYYYYY

Partial derivatives and multidimensional integrals are defined analogously.

\section{Lower and upper Minkowski space-times}

Consider a point $x$ in Minkowski space $\mathbb{R}^4$ and assume
\be
(x^0,x^1,x^2,x^3)=\big(f(X^0),f(X^1),f(X^2),f(X^3)\big)
\ee
where $(x^0,x^1,x^2,x^3)\in \mathbb{R}^4$ and $(X^0,X^1,X^2,X^3)\in \RR^4$. The two spaces are Minkowskian in the sense that the invariant quadratic forms are defined by
\be
g_{ab}x^ax^b
&=&
(x^0)^2-(x^1)^2-(x^2)^2-(x^3)^2,\\
G_{ab}X^aX^b
&=&
(X^0)^{2'}\ominus (X^1)^{2'}\ominus (X^2)^{2'}\ominus (X^3)^{2'}.
\ee
Since
\be
f^{-1}\big(g_{ab}x^ax^b\big)
&=&
f^{-1}\Big(f(X^0)^2-f(X^1)^2-f(X^2)^2
\nonumber\\
&\pp=&
\pp{f^{-1}\Big(}
-f(X^3)^2\Big)\\
&=&
f^{-1}\Big(f(G_{ab})f(X^a)f(X^b)\Big)\\
&=&
G_{ab}X^aX^b,
\ee
the two quadratic forms are related by $f$, and $g_{ab}=f(G_{ab})$. Covariant and contravariant world-vectors are defined in the usual way,
\be
X_a
&=&
G_{ab}X^b\\
&=&
f^{-1}\Big(
\sum_{b=0}^3f(G_{ab})f(X^b)\Big)
\\
&=&
f^{-1}\Big(
\sum_{b=0}^3g_{ab}x^b\Big)=f^{-1}(x_a).
\ee
The quadratic forms are invariant with respect to Lorentz transformations,
\be
x'_a &=& \lambda{_a}{^b}x_b=\sum_{b=0}^3\lambda{_a}{^b}x_b,\label{lambda}\\
X'_a &=& \Lambda{_a}{^b}X_b=\oplus_{b=0}^3\Lambda{_a}{^b}\odot X_b \label{Lambda}\\
&=&
f^{-1}\Big(
\sum_{b=0}^3f(\Lambda{_a}{^b})f(X_b)\Big).
\ee
This type of Lorentz transformation, but for $f$ representing a Cantor set, was explicitly used in \cite{MC,ACK} to construct fractal homogeneous spaces.

We will later need an explicit boost,
\be
\Lambda{^a}{_b}
&=&
\left(
\begin{array}{cccc}
\Cosh\phi & \ominus\Sinh\phi & 0' &0'\\
\ominus\Sinh\phi & \Cosh\phi & 0' &0'\\
0' & 0' & 1' & 0'\\
0' & 0' & 0' & 1'
\end{array}
\right),\label{boost}
\ee
where $\Sinh\phi=f^{-1}\big(\sinh f(\phi)\big)$, $\Cosh\phi=f^{-1}\big(\cosh f(\phi)\big)$ satisfy
\be
\Cosh^{2'}\phi\ominus \Sinh^{2'}\phi=1'.
\ee
The four-velocity
\be
U{^a}
&=&
\left(
\begin{array}{c}
\Cosh\phi\\
\Sinh\phi\\
0' \\
0'
\end{array}
\right),\label{4U}
\ee
is mapped by (\ref{boost}) into $(1',0',0',0')$.

\section{Minkowski space-time $\mathbb{R}_+^4$}

Now let us make the analysis more explicit. Let the Fechner function $f(X)=\mu\ln X+\nu$, $\mu>0$, be the bijection $f:\mathbb{R}_+\to \mathbb{R}$. Accordingly, $\RR=\mathbb{R}_+$.
$f^{-1}(x)=e^{(x-\nu)/\mu}$, and thus $0'=f^{-1}(0)=e^{-\nu/\mu}$, $1'=f^{-1}(1)=e^{(1-\nu)/\mu}$.

\subsection{Arithmetic}

Let us begin with the explicit form of arithmetic operations.
Addition and subtraction explicitly read
\be
X\oplus Y
&=&
f^{-1}\big(f(X)+f(Y)\big)
\nonumber\\
&=&
XY e^{\nu/\mu},\label{X+Y'}\\
X\ominus Y
&=&
f^{-1}\big(f(X)-f(Y)\big)
\nonumber\\
&=&
e^{-\nu/\mu}X/Y.\label{X-Y'}
\ee
The arithmetic operations occurring at the right sides of (\ref{X+Y'}) and (\ref{X-Y'}) are those from $\mathbb{R}$ and not from \RR (the latter occur at the left sides of these formulas). For example,
$X\oplus 0'=X e^{-\nu/\mu} e^{\nu/\mu}=X$.

Note that although $X>0$ in $f(X)=\mu\ln X+\nu$, one nevertheless has a well defined negative number $\ominus X=0'\ominus X=e^{-2\nu/\mu}/X\in \RR=\mathbb{R}_+$, which is positive from the point of view of the arithmetic of $\rr$. Let us cross-check the negativity of $\ominus X$:
\be
\ominus X\oplus X
&=&
(\ominus X)X e^{\nu/\mu}
\\
&=&
(e^{-2\nu/\mu}/X)X e^{\nu/\mu}
\\
&=&
e^{-\nu/\mu}
=0'.
\ee
$(\RR,\oplus)=(\rr_+,\oplus)$ is a group, as opposed to $(\rr_+,+)$. In consequence, the Minkowski space $\rr_+^4$ is invariant under the non-Diophantine Poincar\'e group.

The multiplication in \RR is explicitly given by
\be
X\odot Y
&=&
f^{-1}\big(f(X)f(Y)\big)
\nonumber\\
&=&
e^{\mu\ln X\ln Y+ \nu\ln X+\nu\ln Y+\nu^2/\mu -\nu/\mu},\label{X.Y'}\\
X\oslash Y
&=&
f^{-1}\big(f(X)/f(Y)\big)
\nonumber\\
&=&
e^{(\ln X+\nu/\mu)/(\mu\ln Y+\nu)-\nu/\mu}.\label{X/Y'}
\ee
Again the expressions at the right-hand sides of (\ref{X.Y'}) and (\ref{X/Y'}) involve the arithmetic of $\rr$.

\subsection{Light cone}

The light cone in $\mathbb{R}_+^4$ consists of vectors satisfying
\be
G_{ab}X^aX^b
&=&
f^{-1}\big(f(G_{ab})f(X^a)f(X^b)\big)\\
&=&
f^{-1}\big(g_{ab}f(X^a)f(X^b)\big)\\
&=& 0'=f^{-1}(0)=e^{-\nu/\mu}.
\ee
This is equivalent to $f(X^0)^2=f(X^1)^2+f(X^2)^2+f(X^3)^2$, i.e.
\be
X^0
&=& f^{-1}\Big(\pm \sqrt{f(X^1)^2+f(X^2)^2+f(X^3)^2}\Big)
\\
&=&
e^{\left(\pm \sqrt{(\ln X^1+\nu/\mu)^2+(\ln X^2+\nu/\mu)^2+(\ln X^3+\nu/\mu)^2}-\nu/\mu\right)}
\nonumber\\
\label{X^0}
\ee
Fig.~1 shows the light cone $G_{ab}X^aX^b=0'$ in $1+2$ dimensional Minkowski space $\mathbb{R}_+^3$.

\begin{figure}
\includegraphics[width=4 cm]{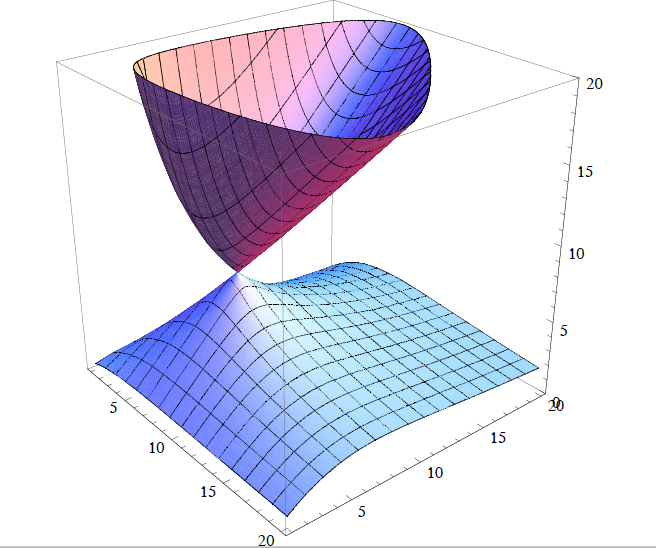}\includegraphics[width=3 cm]{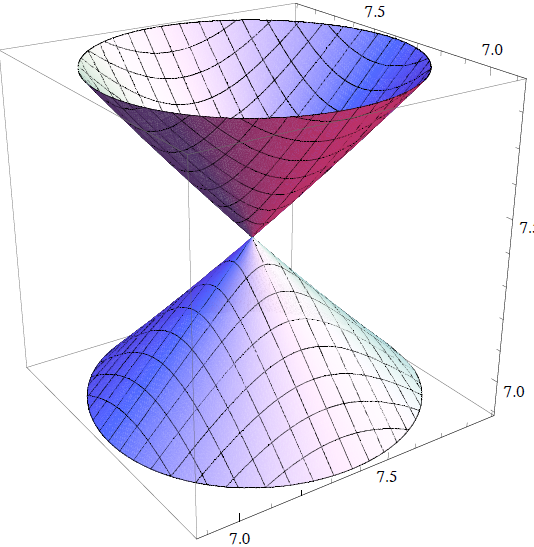}
\caption{Light cone $X_aX^a=0'$ in $1+2$ dimensional Minkowski space $\mathbb{R}_+^3$, for $\mu=10$, $\nu=-20$, and its close-up (right) in the neighborhood  of $(0',0',0')$, where $0'=e^{-\nu/\mu}=e^{2}\approx 7.39$.}
\end{figure}

An arbitrarily located light cone
\be
G_{ab}(X^a\ominus Y^a)(X^b\ominus Y^b)=0'
\ee
corresponds to
\be
g_{ab}\big(f(X^a)-f(Y^a)\big)\big(f(X^b)-f(Y^b)\big)=0,
\ee
that is
\be
X^0=Y^0e^{\pm \sqrt{\ln^2(X^1/Y^1)+\ln^2(X^2/Y^2)+\ln^2(X^3/Y^3)}}.\label{X^0'}
\ee
For $Y^0=\dots=Y^3=0'=e^{-\nu/\mu}$ we reconstruct (\ref{X^0}).

Fig.~2 shows the light cones (\ref{X^0'}) in small neighborhoods of various origins $Y^a$. The plots suggest that a Lorentzian geometry is typical of both the standard `lower-case' space-time $\rr^3$, and of the `upper-case' $\RR^3$. Recall that the latter is also globally Minkowskian, but with respect to the non-Diophantine calculus. Interestingly,
when we employ in $\RR^3$ the miss-matched formalism taken from $\rr^3$, the formulas are locally Lorentzian. The further away from the `walls' of $\mathbb{R}_+^3$ the observation is performed, the more Minkowskian the geometry appears, provided one does not observe objects that are too far away from the observer, as we shall see later.

To prove the local Lorentzian structure analytically, let $X^a=Y^a+\epsilon^a$. Here `$+$' is from $\rr$ since the observer is assumed to perform his analysis in the `wrong' formalism. Then, for $|\epsilon^a/Y^a|\ll 1$,
\be
g_{ab}f'(Y^a)\epsilon^af'(Y^b)\epsilon^b=\tilde g_{ab}(Y)\epsilon^a\epsilon^b\approx 0
\ee
where
\be
\tilde g_{ab}
&=&
g_{ab}f'(Y^a)f'(Y^b)\quad\textrm{(no sum)}\\
&=&
\textrm{diag}\big((Y^0)^{-2},-(Y^1)^{-2},-(Y^2)^{-2},-(Y^3)^{-2}\big)\nonumber\\
\ee
is the Lorentzian metric. $\tilde g_{ab}$ becomes just a conformally rescaled Minkowskian $g_{ab}$ if $Y^0=Y^1=Y^2=Y^3$.

The same effect occurs for general hyperboloids $G_{ab}(X^a\ominus Y^a)(X^b\ominus Y^b)\neq 0'$.
\begin{figure}
\includegraphics[width=8 cm]{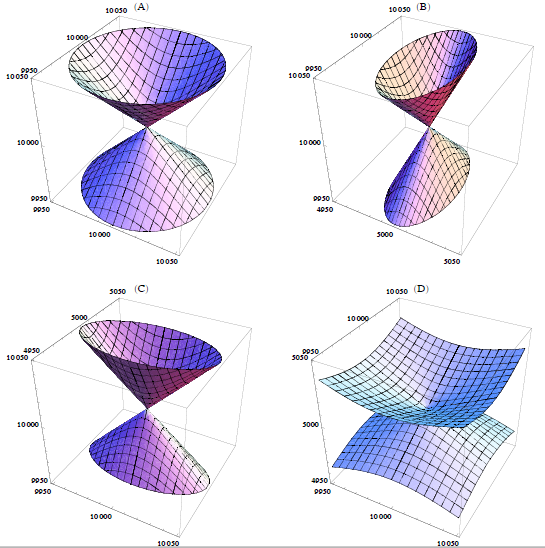}
\caption{Minkowskian correspondence principle in $1+2$ dimensional Fechnerian space-time. Light cone $(X_a\ominus Y_a)(X^a\ominus Y^a)=0'$ for (A) $(Y^0,Y^1,Y^2)=(10000,10001,10002)$, (B) $(Y^0,Y^1,Y^2)=(10000,5000,10000)$,  (C) $(Y^0,Y^1,Y^2)=(10000,10000,5000)$, and (D) $(Y^0,Y^1,Y^2)=(5000,10000,10001)$. The further away from the boundaries of $\mathbb{R}_+^3$, the more Minkowskian-looking the light cones are.}
\end{figure}

\section{Minkowski space-time $(-L/2,L/2)^4$}

Let $\RR=(-L/2,L/2)$ and $f(X)=\tan (\pi X/L)$, $f^{-1}(x)=(L/\pi)\arctan x$. $f:\RR\to \mathbb{R}$ is a possible bijection, with $0'=0$ and $1'=L/4$.
\begin{figure}
\includegraphics[width=4 cm]{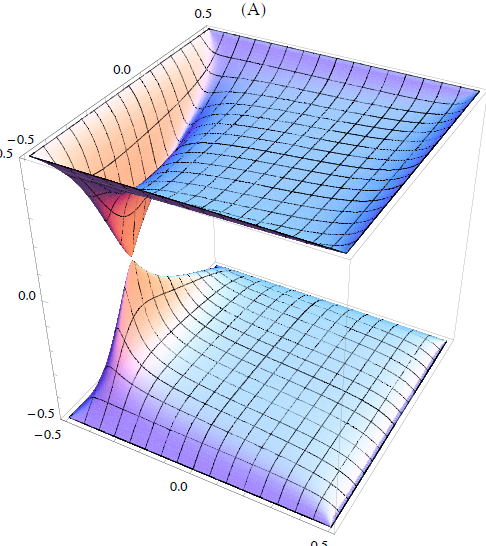}\includegraphics[width=4 cm]{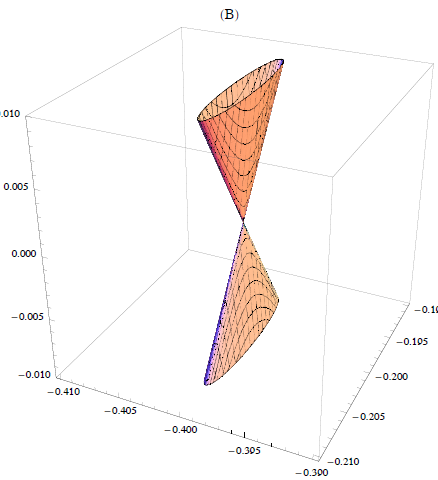}
\caption{Light-cone with the origin at $(Y^0,Y^1,Y^2)=(0,-0.4,-0.2)$ in $1+2$ dimensional Minkowski space $(-0.5,0.5)^3$ ($L=1$). (A) The global picture, and (B) the close-up of the origin of the cone}
\end{figure}
\begin{figure}
\includegraphics[width=4 cm]{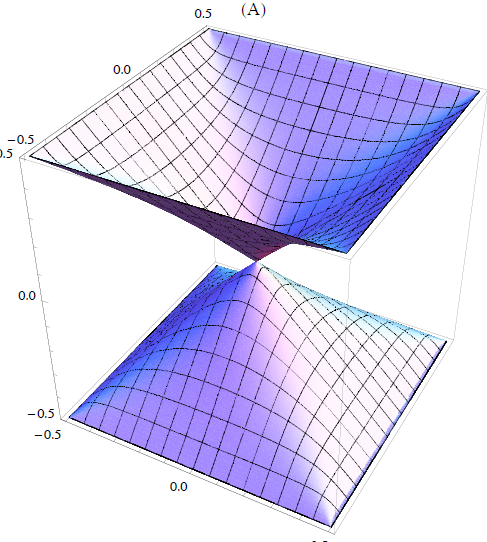}\includegraphics[width=4 cm]{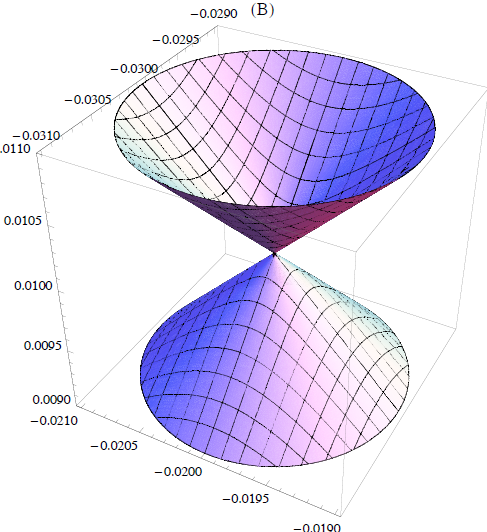}
\caption{The same as in the previous figure, but with $(Y^0,Y^1,Y^2)=(0.01,-0.02,-0.03)$. }
\end{figure}

\begin{widetext}
\subsection{Light cone}

The light cone $G_{ab}(X^a\ominus Y^a)(X^b\ominus Y^b)=0'$ corresponds to
\be
X^0
&=&
f^{-1}\left(f(Y^0)\pm \sqrt{\big(f(X^1)-f(Y^1)\big)^2+\big(f(X^2)-f(Y^2)\big)^2+\big(f(X^3)-f(Y^3)\big)^2}\right)
\\
&=&
\frac{L}{\pi}\arctan\left(\tan\frac{\pi Y^0}{L}
\pm
\sqrt{\left(\tan\frac{\pi X^1}{L}-\tan\frac{\pi Y^1}{L}\right)^2
+
\left(\tan\frac{\pi X^2}{L}-\tan\frac{\pi Y^2}{L}\right)^2
+
\left(\tan\frac{\pi X^3}{L}-\tan\frac{\pi Y^3}{L}\right)^2}\right).
\nonumber\\
\ee
Now let $X^a=Y^a+\epsilon^a$, where $|\epsilon^a/L|\ll 1$. The effective Lorentzian metric in a neighborhood of $Y$ reads
\be
\tilde g_{ab}(Y)
&=&
\frac{\pi^2}{L^2}\,
\textrm{diag}\left(\cos^{-4}\frac{\pi Y^0}{L},-\cos^{-4}\frac{\pi Y^1}{L},-\cos^{-4}\frac{\pi Y^2}{L},-\cos^{-4}\frac{\pi Y^3}{L}\right).
\ee
\end{widetext}
Another example of a bijection $f:\RR\to\mathbb{R}$ is provided by $f(X)=\textrm{arctanh}(2X/L)$, $f^{-1}(x)=(L/2)\tanh x$.

\section{Field produced by a pointlike charge}

The above two examples show that several different types of arithmetic may coexist in the same space-time. Let us try to find a phenomenon where a nontrivial $f$ can be detected.

I propose to concentrate on Maxwell equations, but formulated in terms of this form of arithmetic that makes the geometry globally Minkowskian.
Consider the d'Alembertian
\be
\Box' &=& G_{ab}\frac{D}{DX^a}\frac{D}{DX^b}
\ee
defined with respect to non-Diophantine partial derivatives. Let $\RR\ni S\mapsto X^a(S)\in \RR^4$ be a world-line of a pointlike charge, and let $J^a(X)$ be the current associated with the world-line. The Maxwell equations can be taken in the form
\be
\Box' A^a(X)=J^a(X).
\ee
The procedure of finding $A^a(X)$ is standard \cite{Jackson}, but we only have to take care of appropriate definitions of non-Diophantine arithmetic and calculus. The end result is
\be
A^a(X)
&=&
f^{-1}\left(\frac{f(C)f\big(U^a(Y)\big)}{g_{bc}f\big(U^b(Y)\big)\big(f(X^c)-f(Y^c)\big)}\right).\label{AX1}
\ee
Here $C$ is a constant, $U^a(X)$ is the four-velocity of the charge, and $X^b\ominus Y^b$ is future-pointing and null.
We say that $X^a\in \RR^4$ is future-pointing if $x^a=f(X^a)\in\rr^4$ is future-pointing.
The summation convention is applied unless otherwise stated.

The four-velocity is normalized by $U_aU^a=1'$ (where $1'=f^{-1}(1)$ is the neutral element of multiplication), which is equivalent to
\be
g_{ab}f\big(U^a(Y)\big)f\big(U^b(Y)\big)=1.
\ee

The four-potential is a world-vector gauge field and thus under the action of a Lorentz transformation $\Lambda$ transforms by $A^a(X)\mapsto A'^a(X)=(\Lambda A)^a(\Lambda^{-1}X)$, up to a gauge transformation. In standard notation
\be
A'^a(X) &=& C\frac{\Lambda U^a(Y)}{U_b(Y)\big((\Lambda^{-1}X)^b-Y^b\big)}
\\
&=& C\frac{\Lambda U^a(Y)}{(\Lambda U)_b(Y)\big(X^b-(\Lambda Y)^b\big)}.
\label{AX'}
\ee

Now consider a $Y$-independent $U^a$ of the form (\ref{4U}) and let $\Lambda{^a}{_b}$ be given by (\ref{boost}). To simplify further analysis let us take the point of observation at the origin $X^a=(0',0',0',0')\equiv 0'$ and the source at $Y^a=(Y^0,Y^1,0',0')$. $Y^a$ is null and past-pointing, which implies $f(Y^0)=-|f(Y^1)|$ (notice that in our examples we assume that $f$ is increasing and there exist various arguments why this is important for physical consistency of the formalism). Accordingly,
\be
(\Lambda Y)^0
&=&
Y^0\odot\Cosh\phi\ominus Y^1\odot\Sinh\phi,\\
f\big((\Lambda Y)^0\big)
&=&
-|f(Y^1)|\left(\cosh f(\phi)+\frac{f(Y^1)}{|f(Y^1)|}\sinh f(\phi)\right)\nonumber\\
&=&
-|f(Y^1)|
e^{\frac{f(Y^1)}{|f(Y^1)|}f(\phi)}.
\ee
The only non-vanishing component of the potential reads
\be
A'^0(0')
&=&
C\oslash (\ominus \Lambda Y)^0
\\
&=&
f^{-1}\left(\frac{f(C)}{f(0')-f\big((\Lambda Y)^0\big)}\right)\\
&=&
f^{-1}\left(\frac{f(C)}{|f(Y^1)|
e^{\frac{f(Y^1)}{|f(Y^1)|}f(\phi)}}\right).
\ee
In order to continue we have to make $f$ more concrete. Let us begin with the Minkowski space $(-L/2,L/2)^4$.

\section{Fields in $(-L/2,L/2)^4$}

$f(X)=\tan(\pi X/L)$ implies
\be
A'^0(0')
&=&
\frac{L}{\pi}\arctan\left(\frac{f(C)}{|\tan(\pi Y^1/L)|
e^{\frac{Y^1}{|Y^1|}f(\phi)}}\right).
\ee
Let us first consider the case of $|\pi Y^1/L|\ll 1$, but with $Y^1$ sufficiently far from the singularity at the origin $0'=0$, so that the arguments of both $\tan$ and $\arctan$ are small. The field is then approximately Coulombian
\be
A'^0(0')
&\approx&
\frac{L}{\pi}\frac{f(C)}{|\pi Y^1/L|
e^{\frac{Y^1}{|Y^1|}f(\phi)}}\\
&=&
\frac{L^2}{\pi^2}f(C)\frac{1}{|\bm{Y}|}e^{-\frac{Y^1}{|Y^1|}f(\phi)}.
\ee
Notice that up to this point we have worked with linearized forms of $f$ and $f^{-1}$ since for small distances and small potentials it is enough if one replaces exact nonlinear $f$ by its linear approximation.

Let us concentrate on a charge at rest, i.e. with $f(\phi)=0$, and denote $q=L^2f(C)/\pi^2$. We conclude that a source placed in a neighborhood of the origin produces the Coulomb field whose value at the point of observation is
\be
A'^0(0')
&\approx&
\frac{q}{|\bm{Y}|}
\ee
where $|\bm{Y}|$ is the distance from the source.

Now, let us increase $|\bm{Y}|$ so that $\tan$ cannot be anymore approximated by its argument. The argument of $\arctan$ is even smaller, so here the approximation is still valid. The result is
\be
A'^0(0')
&\approx&
\frac{L}{\pi}\frac{f(C)}{|\tan(\pi Y^1/L)|}
\\
&=&
\frac{q}{|\bm{Y}|}\frac{\pi Y^1/L}{\tan(\pi Y^1/L)}
\\
&=&
\frac{q}{|\bm{Y}|}e^{-\frac{Y^1}{|Y^1|}f\big(\phi(Y^1)\big)}
\ee
So, in spite of our assumption that in the neighborhood of the origin the charge is at rest, at large distances the charge looks like moving with the four-velocity determined by certain $\phi(Y^1)$, defined by
\be
e^{\frac{Y^1}{|Y^1|}f\big(\phi(Y^1)\big)}
&=&
\frac{\tan(\pi Y^1/L)}{\pi Y^1/L}.
\ee
The velocity $\beta=\tanh f\big(\phi(Y^1)\big)$ is
\be
\beta
&=&
\frac{Y^1}{|Y^1|}\frac{\left(\frac{\tan(\pi Y^1/L)}{\pi Y^1/L}\right)^2-1}{\left(\frac{\tan(\pi Y^1/L)}{\pi Y^1/L}\right)^2+1}.\label{beta fin}
\ee
$\beta$ is the velocity deduced by the observer located at $X^a=0'$ who analyses his data on the basis of the `standard' Diophantine arithmetic, whereas the physical non-Diophantine arithmetic, employed in Maxwell's equations, is given by $\odot$, $\oplus$, etc.
The observer is related with the physical Universe by means of an `information channel' $f$, but is unaware of it. The miss-match of mathematical structures leads to unexpected behavior of distant objects.

Had the observer decided to employ Einstein's general relativity, a similar miss-match would have occurred. A solution would produce an expanding universe whose behavior would be consistent with the Hubble law at small distances, but very distant objects would acquire an unexplained acceleration. We will demonstrate this explicitly in section IX.

As our second example consider
\be
A'^0(0')
=
\frac{L}{2}\tanh\left(\frac{f(C)}{|\textrm{arctanh}(2Y^1/L)|e^{\frac{Y^1}{|Y^1|}f(\phi)}}
\right).
\ee
With the same approximations as before
\be
A'^0(0')
&\approx&
\frac{L}{2}\frac{f(C)}{|2Y^1/L|}e^{-\frac{Y^1}{|Y^1|}f(\phi)}
\\
&=&
\frac{q}{|\bm{Y}|}e^{-\frac{Y^1}{|Y^1|}f(\phi)},
\ee
where $q=L^2f(C)/4$ and $Y$ is in a neighborhood of the origin. For $Y$ further away from the origin and with $f(\phi)=0$ we get
\be
A'^0(0')
&\approx&
\frac{L}{2}\frac{f(C)}{|\textrm{arctanh}(2Y^1/L)|}
\\
&=&
\frac{q}{|\bm{Y}|}\frac{2Y^1/L}{\textrm{arctanh}(2Y^1/L)}.
\ee
Repeating the remaining calculations we obtain
\be
\beta
&=&
\frac{Y^1}{|Y^1|}\frac{\left(\frac{\textrm{arctanh}(2 Y^1/L)}{2 Y^1/L}\right)^2-1}{\left(\frac{\textrm{arctanh}(2 Y^1/L)}{2 Y^1/L}\right)^2+1}.
\label{beta fin1}
\ee
\begin{figure}
\includegraphics[width=8 cm]{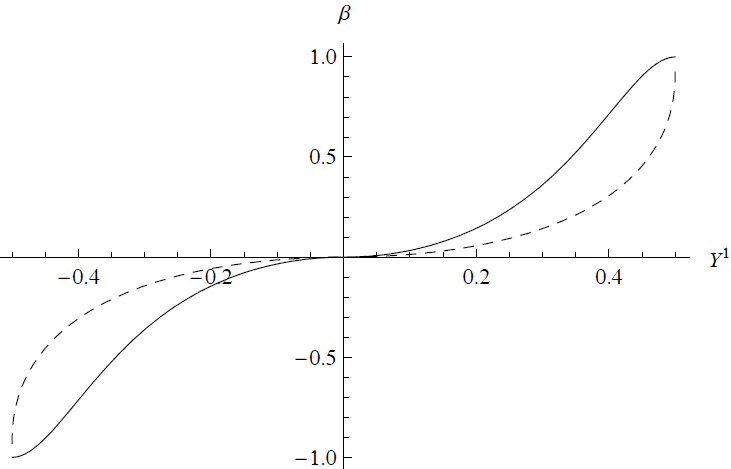}
\caption{The plot of (\ref{beta fin}) (full) and (\ref{beta fin1}) (dashed) for $L=1$. The velocity has the same sign as $Y^1$ so the motion is toward the horizons $Y^1=\pm 1/2$.}
\end{figure}
Fig.~5 compares (\ref{beta fin}) with (\ref{beta fin1}). In both cases the charge looks as if it moved toward the boundaries $Y^1=\pm L/2$, where it approaches the velocity of light. The motion is non-trivially accelerated.

\section{Fields in $\mathbb{R}_+^4$}

It is instructive to consider explicitly also the case of $\mathbb{R}_+^4$ since $0'=e^{-\nu/\mu}$ and thus certain counterintuitive elements of a non-Diophantine arithmetic and calculus become more visible. The example will make further generalizations easier to understand.

Here $f(X)=\mu\ln X+\nu$, $\mu>0$, $f^{-1}(x)=e^{(x-\nu)/\mu}\approx e^{-\nu/\mu}(1+x/\mu)=0'+xe^{-\nu/\mu}/\mu$. Setting $Y^1=0'+r$,
$f(0'+r)=\mu \ln [0'(1+r/0')]+\nu=\mu \ln (1+r/0')$, and assuming $|r|\ll 1$, we obtain
\be
A'^0(0')
&=&
f^{-1}\left(\frac{f(C)}{|f(Y^1)|
e^{\frac{f(Y^1)}{|f(Y^1)|}f(\phi)}}\right)
\\
&\approx&
0'
+
\frac{f(C)e^{-\nu/\mu}/\mu^2}{|\ln(1+r/0')|
e^{\frac{f(Y^1)}{|f(Y^1)|}f(\phi)}}
\\
&\approx&
0'
+
\frac{f(C)(e^{-\nu/\mu}/\mu)^2}{|r|e^{\frac{r}{|r|}f(\phi)}}.
\ee
We identify $q=f(C)(e^{-\nu/\mu}/\mu)^2$. For $f(\phi)=0$ and larger $r$
\be
A'^0(0')
&\approx&
0'
+
\frac{f(C)e^{-\nu/\mu}/\mu^2}{|\ln(1+r/0')|}
\\
&=&
0'
+
\frac{q}{|r|\left|\frac{\ln(1+r/0')}{r/0'}\right|}.
\ee
An observer located at $0'$ will conclude that the field is produced by a charge which satisfies
\be
e^{\frac{r}{|r|}f(\phi)}
&=&
\left|\frac{\ln(1+r/0')}{r/0'}\right|,
\ee
hence
\be
\beta
&=&
\frac{r}{|r|}\frac{\left(\frac{\ln(1+r/0')}{r/0'}\right)^2-1}{\left(\frac{\ln(1+r/0')}{r/0'}\right)^2+1}.\label{beta fin4}
\ee
Fig.~6 shows that $\beta$ given by (\ref{beta fin4}) is always negative: The charge moves toward the horizon $Y^1=0$. Fields produced by charges located between the observer and the horizon would be red-shifted. However, an observer located between the horizon and the charge would detect a blue-shifted field.
\begin{figure}
\includegraphics[width=8 cm]{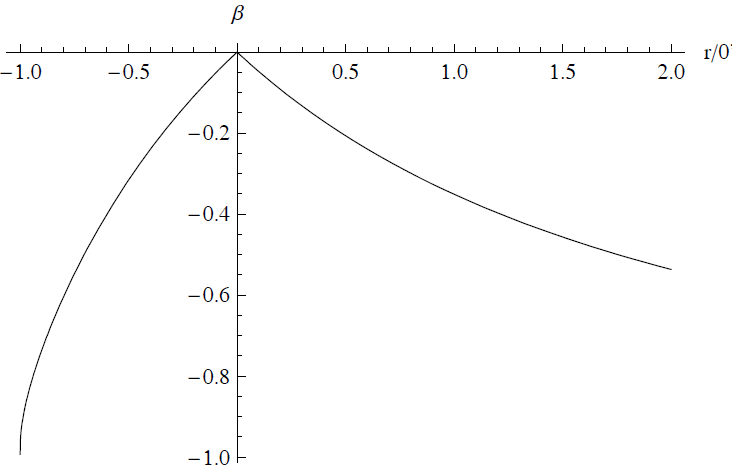}
\caption{The plot of  (\ref{beta fin4}): $\beta$ as a function of $r/0'$.}
\end{figure}

\section{Friedman equation}

Taylor expansions of (\ref{beta fin}), (\ref{beta fin1}) in a neighborhood of $Y^1=0$ begin with third order terms $\sim(Y^1/L)^3$. The effect is small. However, when we switch to non-Diophantine generalized Einstein equations, the correction should become visible at large distances.

So, let us consider the Friedman equation for a flat, matter dominated FRW model with exactly vanishing cosmological constant \cite{Hartle}.
In matter dominated cosmology ($\Omega_m=1,\Omega_r=0$), with no dark energy ($\Omega_v=0$), the scale factor is given by $a(t)=(t/t_0)^{2/3}$. In the non-Diophantine notation the solution reads
\be
A(T)
&=&
(T\oslash T_0)^{2'\oslash 3'}
\\
&=&
f^{-1}\left(
\big(f(T)/f(T_0)\big)^{2/3}\right). \label{A(T)}
\ee
To make (\ref{A(T)}) more explicit, let us experiment with some $f$. For example, the choice of $\RR=(-L/2,L/2)$ and $f(X)=\tan (\pi X/L)$, $f^{-1}(x)=(L/\pi)\arctan x$ leads to the scale factor depicted at Fig.~7, where (\ref{A(T)}) is compared with the standard $(t/t_0)^{2/3}$, for $L=20$, $T_0=1'$, and $t_0$ chosen in a way guaranteeing a reasonable fit of the two plots.
\begin{figure}
\includegraphics[width=8 cm]{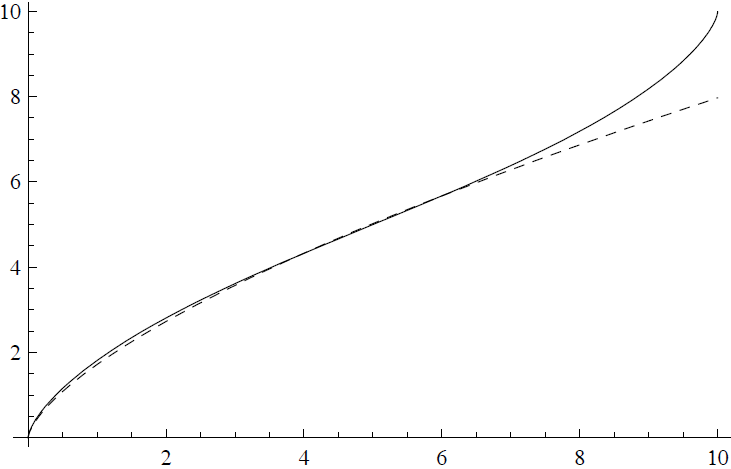}
\caption{The plot of  (\ref{A(T)}) (full) as compared with $(t/t_0)^{2/3}$ for $f(X)=\tan (\pi X/L)$. Both models involve no dark energy, i.e. $\Omega_v=0=0'$.}
\end{figure}

The curve bends up in a characteristic way, typical of dark-energy models of accelerating Universe.
The effect is of purely arithmetic origin, with no need of dark energy. Arithmetic becomes as physical as geometry.

\section{Conclusions}

Gravity is geometry. Is dark energy just arithmetic? Is dark energy dark, because it is always darkest under the lantern, and we are so accustomed to `plus' and `times' that we overlooked the fundamental ambiguity of these operations? Is there a physical law that determines the form of arithmetic, a kind of Einstein equation for $f$?

The questions are relatively well posed. We can change the paradigm and do physics with unspecified $f$, leaving determination of $f$ to experimentalists. Perhaps one day we will understand which arithmetic is physical, and why.

\end{document}